\documentclass{article}

\usepackage{PRIMEArXiv}
\usepackage{algpseudocode}
\usepackage{algorithm}
\usepackage{xcolor}     
\usepackage{amsmath}    
\usepackage[utf8]{inputenc} 
\usepackage[T1]{fontenc}    
\usepackage{hyperref}       
\usepackage{url}            
\usepackage{booktabs}       
\usepackage{amsfonts}       
\usepackage{nicefrac}       
\usepackage{microtype}      
\usepackage{lipsum}
\usepackage{fancyhdr}       
\usepackage{graphicx}       
\graphicspath{{media/}}     
\usepackage{accents}

\newcommand{\bs}[1]{\boldsymbol{#1}}

\title{Can Transformers overcome the lack of data in the simulation of history-dependent flows?
}

\author{
  P. Urdeitx$^{1}$, I. Alfaro$^{1}$, D. Gonz\'alez$^{1}$, F. Chinesta$^{2,3}$, E. Cueto$^{1}$  \vspace{0.5cm}\\
  $^{1}$ ESI Group-UZ Chair of the National Strategy on Artificial Intelligence. \protect \\Aragon Institute of Engineering Research (I3A). Universidad de Zaragoza. Zaragoza, Spain. \\
  $^{2}$ ESI Group Chair. PIMM lab., ENSAM Arts et Métiers Institute of Technology, Paris, France. \\
$^3$ CNRS@CREATE LTD. Singapore.    \\
}

\begin{document}
\maketitle

\begin{abstract}

It is well known that the lack of information about certain variables necessary for the description of a dynamical system leads to the introduction of historical dependence (lack of Markovian character of the model) and noise. Traditionally, scientists have made up for these shortcomings by designing phenomenological variables that take into account this historical dependence (typically, conformational tensors in fluids).

Often these phenomenological variables are not easily measurable experimentally. In this work we study to what extent Transformer architectures are able to cope with the lack of experimental data on these variables.

The methodology is evaluated on three benchmark problems: a cylinder flow with no history dependence, a viscoelastic Couette flow modeled via the Oldroyd-B formalism, and a non-linear polymeric fluid described by the FENE model. Our results show that the Transformer outperforms a thermodynamically consistent, structure-preserving neural network with metriplectic bias in systems with missing experimental data, providing lower errors even in low-dimensional latent spaces. In contrast, for systems whose state variables can be fully known, the metriplectic model achieves superior performance.

\end{abstract}


\section{Introduction}


The modeling of complex physical systems has recently evolved toward a new paradigm that combines classical physics-based approaches with modern data-driven modeling techniques \cite{Koumoutsakos2025}. Traditional mechanistic models based on physics make it possible to identify the fundamental laws that allow the behaviour of systems to be extrapolated. However, these models require precise characterization of the system's state variables, which is not always feasible, as real-world systems often involve inaccessible variables—quantities that cannot be directly measured or tracked. On the other hand, recent advances in data-driven models, fueled by the development of deep neural networks, have demonstrated remarkable capabilities in capturing complex patterns across diverse domains. While their effectiveness is undeniable in fields like computer vision and natural language processing, where no predefined functions or governing differential equations exist, these models still face significant limitations in interpretability and out-of-distribution generalization.

To leverage the strengths of both paradigms, the development of hybrid architectures is emerging as a promising direction, integrating physical and geometric priors into neural network frameworks under the emerging discipline of physics-enhanced machine learning \cite{Cicirello2024, Cuomo2022}. These models encode physical knowledge through inductive or learning biases that structure the learning process and enhance generalization \cite{Faroughi2022}. This approach has proven particularly valuable in physical systems where conservation laws, symmetries, or thermodynamic constraints provide rigorous guidance to restrict the model's hypothesis space \cite{Hernandez2021a}.

In general, the construction of a model of a given physical phenomenon, whether mechanistic or data-driven, requires the choice of a set of state variables that allow the characterisation of the energy of the system \cite{Grmela1997}. When this description becomes too complicated (think of the position and momenta of particles in a description based on molecular dynamics, which requires controlling the Avogadro number of particles to obtain meaningful statistics), the scientist has to choose certain variables that will not be resolved. When the fully solved variables are projected to a scale where certain details are omitted, the so-called Generalised Langevin Equation (GLE) is obtained \cite{Langevin1,Lei14183,Langevin3,Langevin4}.

To obtain the GLE, consider, for simplicity, a system whose dynamics is described by only two degrees of freedom, see \cite{gonzalez2021learning,ma2018model,chinesta2020learning}, 
\begin{subequations}\label{degeneracy}
\begin{alignat}{1}
\dot{x} &= A_{11}x+A_{12}y,\label{a1}\\
\dot{y} &= A_{21}x+A_{22}y, \label{a2}
\end{alignat}
\end{subequations}
with $x(0)=x_0$ and $y(0)=y_0$. Suppose now that for some reason we cannot have access to the measurement of the variable $y$. If we substitute $y$ into Eq. (\ref{a1}), we obtain an evolution equation for the only variable we can measure, $x$,
\begin{equation}\label{GLE}
\dot{x} = \underbrace{A_{11}x}_{\text{Markovian}} + \underbrace{A_{12}\int_0^t e^{A_{22}(t-s)} A_{21}x(s)ds}_{\text{Non-Markovian}} + \underbrace{A_{12}e^{A_{22}t}y_0}_{\text{Noise}}.
\end{equation}
The first term is Markovian, but the second one is non-Markovian (depends on history) and the third one is equivalent to noise, since it depends solely on $y_0$, the initial condition of a variable that we can not measure nor control. Eq.~(\ref{GLE}) is the well-known Mori-Zwanzig formalism for a linear case \cite{Mori,Zwanzig}.

This difficulty has traditionally been solved by means of phenomenological variables, the experimental measurement of which is, however, extremely complicated. These approaches leverage knowledge of past events to define their current state and predict their evolution. In viscoelastic fluids, the evolution depends not only on the instantaneous state but also on past trajectories encoded in phenomenological variables like the conformation tensor \cite{owens2002computational,Binns2024}. These phenomenological variables translate microscopic transformations, such as polymer fiber alignment or crystalline lattice dislocations, into meso/macroscopic descriptors for traditional modeling.

In this context, history-dependent systems pose a major challenge in building data-driven models \cite{ayensa2018new}. Hybrid methodologies, which combine data-driven models with physics-based approaches, struggle in addressing this complexity \cite{Sitapure2023, Liu2025}. Even thermodynamically consistent techniques, see \cite{gonzalez2019thermodynamically,cueto2023thermodynamics}, in the absence of experimental data on certain state variables, have difficulties with data-driven modelling, as will be seen below.

To try to solve these difficulties, several authors have proposed scientific machine learning techniques that include architectures that, in a way, take history into account \cite{funahashi1993approximation,kimura1998learning, Yu2019,rajendra2020modeling,wang2017new}. These include Recurrent Neural Networks, RNN, or Long Short-Term Memory architectures, among others. Similarly, convolutional neural networks (CNNs) have been applied to extract historical information from system data \cite{Pandey2019, Lara-Bentez2020}. As with RNNs, the accessible temporal window in CNNs is restricted and computationally expensive to expand.  However, in the field of large language models, an architecture that stands out for its ability to put context to time series (of words, in this case) has gained enormous popularity and has aroused great curiosity about its possible behaviour in predicting the behaviour of physical systems. A transformer is a neural network architecture built around the multi-head attention mechanism. It converts text into numerical representations known as tokens, and at each layer, these tokens are contextualized based on the information available within the context window \cite{Vaswani2017}. The use of transformers for the prediction of the behaviour of physical systems has however a much more limited tradition \cite{Geneva2022}.

Transformers present two key qualities that make them particularly suitable for dynamic system modeling \cite{Alkin2024, Cao2021}. The first and most critical advantage is their ability to capture non-linear dependencies across distant time steps, making them highly effective for modeling materials with complex, history-dependent behavior. The second, equally important quality is their non-sequential integration of system evolution, which bypasses stepwise computation and enables parallelized prediction reconstruction, resulting in significantly improved computational efficiency \cite{Fournier2023}.

Transformers operate through purely data-driven approximations without incorporating explicit physical knowledge \cite{Vaswani2017}. However, recent work has demonstrated how integrating physical biases into these architectures substantially enhances their predictive performance and reliability. Notable examples include implementations that embed Koopman operators to linearize nonlinear dynamics \cite{Geneva2022}, symplectic transformers that preserve Hamiltonian invariants \cite{Brantner2023}, and methods that enforce pointwise satisfaction of partial differential equations through regularized loss terms \cite{Shih2024}.

While these approaches advance the field of physics-enhanced machine learning, their scope remains limited to local constraints, which fail to ensure global thermodynamic principles such as energy conservation and irreversibility. To address this methodological gap, recent studies have proposed architectures based on Hamiltonian (for conservative systems) or metriplectic formalisms (for dissipative phenomena)\cite{gonzalez2019thermodynamically,cueto2023thermodynamics, Hernandez2021, Hernandez2021a, Zhang2022,Gruber2024}. These formalisms enable unveiling a thermodynamically consistent latent space while ensuring the fulfillment of the laws of thermodynamics (energy conservation in closed systems, non-negative entropy production). However, paradoxically, their application faces a fundamental challenge: the need to define explicit phenomenological variables storing the full historical knowledge of the system, precisely the missing data in non-Markovian temporal dependence contexts. 

In order to study to what extent Transformers can help to alleviate the dependence of the models on the aforementioned phenomenological variables, this paper makes a comparison between a thermodynamically consistent architecture, which will be considered as the baseline methodology, and a Transformer-based architecture operating on a thermodynamically consistent latent space. The thermodynamically consistent architecture has been chosen because it has been shown to provide results that improve the accuracy of black-box techniques based on the use of neural networks. Of course, there is a plethora of different techniques that could have been compared, but thermodynamic consistency is a quality that comes in handy in this comparison, given that the problem stems from an incomplete description of the energy of the system.

On the above mentioned thermodynamically consistent latent space, two time integration schemes will be compared: one based on a metriplectic formalism and the other one based on the use of a Transformer. Both methodologies are compared in problems without historical dependence, for which there is no need of phenomenological variables (i.e., Navier-Stokes) and in problems that require the knowledge of microscopic variables, both linear and nonlinear. These variables are not experimentally measurable and will therefore be assumed unknown. As will be seen in the numerical examples, the Transformer-based architecture is able to compensate for this lack of knowledge and to provide more accurate results than the metriplectic architecture.

 We propose a sequential paradigm for modeling history-dependent systems: (1) training an encoder-decoder network with a metriplectic integrator to enforce thermodynamic consistency in the latent space \cite{Morrison2009, Grmela2010}; (2) replacing the metriplectic integrator with a transformer that processes temporal sequences while keeping the encoder-decoder frozen \cite{Geneva2022}. In this sense, the self-attention mechanism compensates for missing phenomenological variables by implicitly reconstructing historical dependencies, effectively emulating microstructural evolution (e.g., polymer conformation) through contextual patterns in observable sequences. This autonomous discovery of temporal patterns positions transformers as ideal tools for systems with historical dependence where phenomenological variables lack operational definitions.  The sequential approach enables direct comparison between thermodynamics-driven (metriplectic) and history-driven (Transformer) integration paradigms for the same latent manifold. We validate this on systems including viscoelastic Couette flows, where deliberate omission of the conformation tensor tests the transformer's ability to salvage incomplete physics.

\section{Methods}

This section presents our methodology for modeling the evolution of diverse dynamical systems. Our approach integrates a thermodynamic framework based on metriplectic dynamics with a deep learning strategy, with particular interest in history-dependent physical systems. The framework comprises three key stages:

\begin{enumerate}
    \item[(\romannumeral 1)] Identifying the metriplectic structure of the system: unveiling the underlying metriplectic formalism governing the system's dynamics, assumed to be dissipative.
    \item[(\romannumeral 2)] Encoding the structure into a latent space: constructing a neural architecture to embed this metriplectic structure into a latent representation. This architecture will be considered the reference architecture, as it is able to guarantee thermodynamic consistency and has been shown in the literature to outperform  other existing architectures \cite{Hernandez2021a, Gruber2024, Zhang2022}.
    \item[(\romannumeral 3)] Integrating the latent dynamics: Evolving the system using either a metriplectic integrator or a Transformer architecture trained on sequential data.
\end{enumerate}

The two architectures are compared, taking the metriplectic architecture as a baseline, since it satisfies the laws of thermodynamics by construction. We do it on problems with and without historical dependency, to discern to what extent the Transformer-based architecture is able to alleviate the limitations that the metriplectic formulation presents when phenomenological variables cannot be measured. Each stage is detailed in the following subsections.

\subsection{Metriplectic architecture}

Metriplectic systems describe the evolution of physical systems by coupling reversible and irreversible dynamics through geometric structures: symplectic and metric. The reversible part of the dynamics is represented by the symplectic structure of the manifold, which induces conservative Hamiltonian evolution \cite{Morrison1984, Eldred2023}. In contrast, the dissipative dynamics are modeled through a symmetric, positive semi-definite metric structure, responsible for entropy production and free energy dissipation in the system. This methology can be interpreted as a natural coarse-graining procedure, in which the microscopic description of the system is taken into account with the variational of macroscopic variables. Based on this description, Öttinger and Grmela presented the GENERIC formalism (General Equation for Non-Equilibrium Reversible-Irreversible Coupling) as a general framework to describe thermodynamically consistent dynamics by combining Hamiltonian and dissipative contributions in a unified structure \cite{Grmela1997, Beris2001}. 

Let $ \boldsymbol{z} \in \mathbb{R}^{\tt n}$ denote the state vector of the system. The GENERIC evolution equation is expressed as:
\begin{equation}
\frac{d\boldsymbol{z}}{dt} = \boldsymbol{L}(\boldsymbol{z}) \frac{\partial E}{\partial \boldsymbol{z}} + \boldsymbol{M}(\boldsymbol{z}) \frac{\partial S}{\partial \boldsymbol{z}},
\end{equation}
where $ E(\boldsymbol{z}) $ is the total energy, $ S(\boldsymbol{z}) $ is the entropy, $ \boldsymbol{L}(\boldsymbol{z}) $ is a skew-symmetric operator encoding reversible dynamics (via a Poisson bracket), and $ \boldsymbol{M}(\boldsymbol{z}) $ is a symmetric positive semi-definite operator accounting for irreversible effects (via a dissipative bracket). To ensure thermodynamic consistency, the operators must satisfy the degeneracy conditions:
\begin{equation}
\boldsymbol{M}(\boldsymbol{z}) \frac{\partial E}{\partial \boldsymbol{z}} = \boldsymbol{0}, \qquad \boldsymbol{L}(\boldsymbol{z}) \frac{\partial S}{\partial \boldsymbol{z}} = \boldsymbol{0}.
\end{equation}
These constraints enforce energy conservation and entropy production, respectively, aligning the dynamics with the first and second laws of thermodynamics.

In practical applications, the time integration of such systems can be written in discrete form using a first-order explicit scheme as:
\begin{equation}\label{integrationStep}
\boldsymbol{z}(t+\Delta t) = \boldsymbol{z}(t) + \Delta t  \left[ \boldsymbol{L}(\boldsymbol{z}) \frac{\partial E}{\partial \boldsymbol{z}}(\boldsymbol{z}) + \boldsymbol{M}(\boldsymbol{z}) \frac{\partial S}{\partial \boldsymbol{z}}(\boldsymbol{z}) \right].
\end{equation}

Unlike canonical Hamiltonian systems—defined over position-momentum variables—the GENERIC formalism admits a non-canonical Hamiltonian structure \cite{Morrison1986}, in which the state vector $\boldsymbol{z}$ can consist of macroscopic or transformed observables \cite{Grmela2010}. This generalization is essential for modeling systems where the physical description involves internal or phenomenological variables such as conformation tensors in viscoelastic flows \cite{Laso1993}. These variables are often critical to define the total energy $ E(\boldsymbol{z}) $, yet may be experimentally inaccessible.

In our framework, we employ an encoder-decoder structure to reduce the dimensionality of the system state while preserving sufficient information to reconstruct the physical observables. Although such transformations may, a priori, raise concerns regarding the physical interpretability of the latent space, the non-canonical Hamiltonian formalism provides a geometric justification for applying metriplectic dynamics in transformed coordinates. Specifically, the work of Morrison and Eliezer \cite{Morrison1986} shows that the preservation of structure can be maintained under suitable transformations, legitimizing the use of GENERIC in a learned latent manifold, provided that the reduced variables retain a one-to-one correspondence with the underlying energetic structure of the system.

\subsection{Structure-preserving neural networks}

\begin{figure} 
    \centering
    \includegraphics[width=0.9\textwidth]{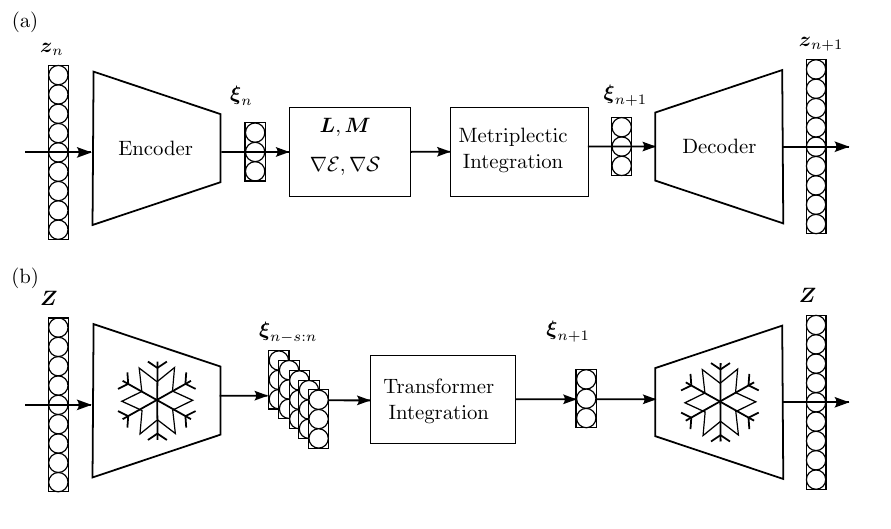}
    \caption{Integration scheme for (a) the Metriplectic Neural Network, and (b) the Transformer-based model. In both cases, the physical state is first mapped to a latent representation via an encoder. In (a), the evolution is computed through a learned metriplectic formalism based on the GENERIC structure, and then decoded back to physical space. In (b), the model receives as input a context window of physical states $\boldsymbol{Z} = \{ \boldsymbol{z}_{n-s}, \dots, \boldsymbol{z}_{n} \}$, which is encoded and passed to the Transformer integrator to predict the next latent state $\boldsymbol{\xi}_{n+1}$. This is decoded to recover the next physical state $\boldsymbol{z}_{n+1}$.}
    \label{Integration}
\end{figure}

Once the metriplectic structure has been defined, we aim to embed it into a neural framework capable of evolving the system's latent representation in a thermodynamically consistent manner. In the standard structure-preserving neural networks (SPNN), the input of the network is formed by the state vector of the system, $\boldsymbol{z}(\bs x,t)$, while the output of the net is taken as the parameters necessary for the reconstruction of the metriplectic formalism: the operators $\bs{L}$, $ \bs{M}$, and the gradients $\nabla E$, $\nabla S$ \cite{Hernandez2021, Urdeitx2024}. In this approach, we defined a metriplectic neural integrator that operates within the reduced latent space, $\boldsymbol{\xi}(t)$, obtained via an encoder-decoder architecture \cite{Hernandez2021a}. The encoder and decoder are responsible for mapping physical observables to and from the latent manifold as:
\begin{align}
\varepsilon_{\phi}:\mathbb{R}^D \rightarrow \mathbb{R}^d, \boldsymbol{\xi}(t)=\varepsilon_{\phi}(\boldsymbol{z}(\boldsymbol{x},t)), \\ 
\delta_{\phi}:\mathbb{R}^d \rightarrow \mathbb{R}^D, \boldsymbol{z}(\boldsymbol{x},t)=\delta_{\phi}(\boldsymbol{\xi}(t).
\end{align}

This latent space is used to learn a metriplectic structure, ensuring consistency with the GENERIC formalism, embedding this way a physical meaning to the latent manifold \cite{Geneva2022}. The encoder-decoder architecture depends on the particular example evaluated. Thus, the main details will be detailed in the results section for each example. 

The system evolution through time steps is then obtained by the integration with the reconstructed formalism, the current state of variables, $\bs \xi_n = \bs \xi(t=n\Delta t)$, and a fixed time step increment, $\Delta t$, leading to the next state $\boldsymbol{\xi}_{n+1}$. The corresponding physical state $\boldsymbol{z}_{n+1} = \boldsymbol{z}(\boldsymbol{x}, t + \Delta t)$ is then reconstructed via the decoder. This integration follows a first-order explicit scheme, as defined in Eq. (\ref{integrationStep}). The overall architecture and information flow are summarized in Fig. \ref{Integration} (a), where the encoder compresses the physical input, $\boldsymbol{z}_n$, to a latent space, $\boldsymbol{\xi}_n$. Then a Feed Forward Neural Network, with configurable depth and units, learns the operators $\bs{L}$, $\bs{M}$, and the gradients $\nabla E$, $\nabla S$. In general, each mapping can be expressed as:
$$
\mathbb{R}^{\tt n_{\text{embd}}} \to \underbrace{\mathbb{R}^{\tt n_{\text{int}}} \to \ldots \to \mathbb{R}^{\tt n_{\text{int}}}}_{\tt n_{\text{layers}} - 1 \text{ hidden layers}} \to \mathbb{R}^{\tt n_{\text{out}}},
$$
being, $\tt n_{\text{embd}}$, the latent space dimension, $\tt n_{\text{int}}$ the dimension of the intermediate hidden layers, and $\tt n_{\text{out}}$ the dimension of the output layer, depending on the specific output ($\bs{L}$, $\bs{M}$ operators, or $\nabla E$, $\nabla S$ gradient vectors). The depth of the intermediate layers is defined by the $\tt n_{\text{layers}}$ hyperparameter. 

$\bs{L}$, the symplectic matrix, is well-known to be skew-symmetric. To ensure this property is satisfied by construction, we parametrize it as $\bs L=\bs l-\bs l^T$, where $\boldsymbol{l}$ is a learnable  matrix ($n_{\text{out}} = (n_{\text{embd}} - 1) n_{\text{embd}}/2$). Similarly, $\bs{M}$, the metric operator, is symmetric and positive semi-definite by construction. To enforce symmetry, it is convenient to learn a matrix $\boldsymbol{m}$ such that $\bs{M} = (\boldsymbol{m} + \boldsymbol{m}^T)$ ($n_{\text{out}} = (n_{\text{embd}} + 1) n_{\text{embd}}/2$) \cite{Hernandez2021}. Finally, $\nabla E$, $\nabla S$ gradients are obteined with $n_{\text{out}}=n_{\text{embd}}$. Then, the latent dynamics are obtained via metriplectic integration, $\boldsymbol{\xi}_{n+1}$, and the result is decoded back to physical space, $\boldsymbol{z}_{n+1}$, to compute the training loss. 

\begin{figure} 
    \centering
    \includegraphics[width=0.95\textwidth]{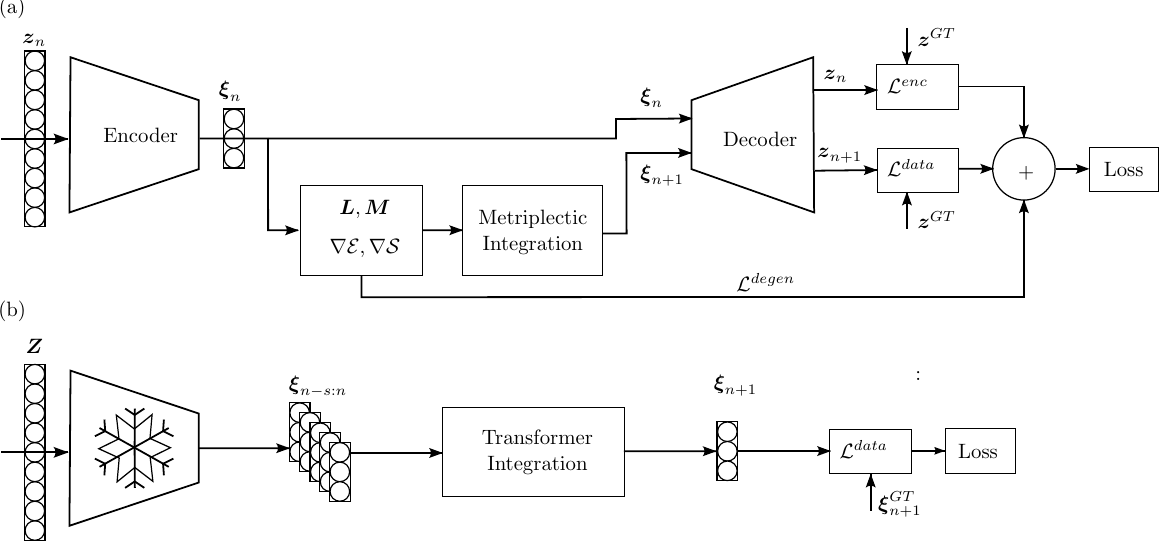}
    \caption{Training scheme for (a) the Metriplectic neural network, and (b) the Transformer-based architecture. In both cases, the physical state $\boldsymbol{z} \in \mathbb{R}^{D}$ is first encoded into a low-dimensional latent representation $\boldsymbol{\xi} \in \mathbb{R}^{d}$. The integration stage then governs the system evolution based on (a) the metriplectic formalism or (b) a context-based Transformer. In the latter case, since the integrator operates on a sequence of latent vectors, the input is derived from a window of physical states $\boldsymbol{Z} = \{ \boldsymbol{z}_{n-s}, \dots, \boldsymbol{z}_{n} \}$.}
    \label{Training}
\end{figure}

It is worth noting, at this point, that the architecture used here is not the same as the one used in the original work on structure-preserving neural networks \cite{Hernandez2021,Hernandez2021a}. In that case, an autoencoder was first trained to find a latent space. This autoencoder was then split in two (encoder and decoder) and in between a metriplectic biased network was trained.
Here, on the other hand, a network comprising an encoder, a metriplectic network and a decoder, which take as input values at time instant $t$ and return values at $t+\Delta t$, is trained at once and monolithically. The latent manifold is intended to have as much thermodynamic information as possible.

The loss function includes three main terms: (i) the reconstruction error of the encoder-decoder pair at time $t$, 
\begin{equation}
    \mathcal{L}^{\text{enc}}_n = \parallel \boldsymbol{z}_{n}^{\text{GT}} - \boldsymbol{z}_{n}^{\text{enc}} \parallel^2_2 .
\end{equation}

(ii) the prediction error after one integration step, 
\begin{equation}
    \mathcal{L}^{\text{data}}_n = \parallel \boldsymbol{z}_{n+1}^{\text{GT}} - \boldsymbol{z}_{n+1}^{\text{net}} \parallel^2_2 .
\end{equation}

and (iii) a degeneracy penalty to softly enforce the GENERIC degeneracy conditions. 
\begin{equation}
    \mathcal{L}^{\text{degen}}_n = \parallel \boldsymbol{L} \nabla S \parallel^2_2 + \parallel \boldsymbol{M} \nabla E \parallel^2_2.
\end{equation}

Then, the global loss function is the sum of the contributions of the loss functions just considered. Due to the differences in the magnitude of each term in the loss, compensation weights should be considered as:
\begin{equation}
\mathcal{L} = \sum_{n=1}^{N_T} \left( \lambda_{\text{enc}} \mathcal{L}^{\text{enc}}_n + \lambda_{\text{data}} \mathcal{L}^{\text{data}}_n + \lambda_{\text{deg}} \mathcal{L}^{\text{degen}}_n \right).  
\end{equation}
The $\lambda_{\text{enc}}$, and $\lambda_{\text{deg}}$, were the encoder-decoder and degeneration weight compensation hyperparameters, respectively, and the weight associated to the prediction error is taken unbalanced, $\lambda_{\text{data}}=1$. $N_T$ represents the number of snapshots in each simulation. Regularization of the network is imposed by using the Adam optimizer during training. Together, these components guide the learning of a themrodynamically consistent latent evolution. The training scheme of the SPNN is represented in Fig. \ref{Training} (a).

\subsection{Transformer}

The transformer architecture was originally developed for natural language processing (NLP) tasks, where it demonstrated an exceptional capacity to model long-range dependencies between tokens in a sentence \cite{Vaswani2017}. Its core mechanism, self-attention, constructs a contextual representation by correlating elements of an input sequence through a position-invariant mapping defined by a learned-attention kernel. The attention window determines the effective temporal or structural correlation range and can be tuned to capture nonlocal interactions.

Beyond its success in NLP, this attention-based paradigm is particularly appealing for modeling physical systems governed by temporal dynamics \cite{Li2022, Wen2022, Geneva2022}. In this work, we focus on dynamic systems whose evolution is influenced by internal variables with memory effects. We aim to evaluate whether, in the absence of explicit access to such variables, the attention mechanism can infer and encode the required temporal dependencies directly from the observable history. The underlying hypothesis is that phenomenological variables may become implicitly encoded in the attention-weighted latent context built across the sequence.

Despite the contextual capacity of transformers, their latent embeddings typically lack direct physical interpretability. To address this issue, several authors have proposed the introduction of physics-based constraints during latent space construction, such as forcing the latent dynamics to comply with known integration formalisms \cite{Geneva2022, Brantner2023}. In our case, we first impose a metriplectic structure on the latent evolution by training an encoder-decoder architecture in conjunction with a metriplectic neural integrator. This leads to a latent manifold $\boldsymbol{\xi}(t)$ that embeds thermodynamic consistency, grounded in the GENERIC formalism. Once this manifold is trained, the encoder and decoder networks are frozen, and the metriplectic integrator is replaced by a transformer model. 
The transformer receives as input a sequence of latent states $(\boldsymbol{\xi}_n, \boldsymbol{\xi}_{n-1}, \dots, \boldsymbol{\xi}_{n-s})$, and returns a prediction of the next latent state $\boldsymbol{\xi}_{n+1}$, following an autoregressive approach. The output is then mapped to the physical variables via the frozen decoder: $\hat{\boldsymbol{z}}_{n+1} = \delta_\phi(\boldsymbol{\xi}_{n+1})$, as illustrated in Fig. \ref{Integration} (b). 

The transformer is trained to reproduce the time evolution of the latent representation inferred by the encoder, given a fixed historical context. Thus, the loss function, $\mathcal{L}_{\text{trans}}$, is defined based on the mean squared error (MSE) between the predicted and ground-truth latent states. More formally, the loss function is defined as:
\begin{equation}
\mathcal{L}_{\text{trans}}^{\text{data}} = \left\| \boldsymbol{\xi}_{n+1}^{\text{GT}} - \boldsymbol{\xi}_{n+1}^{\text{net}} \right\|_2^2,
\end{equation}
where $\boldsymbol{\xi}_{n+1}^{\text{GT}}$, and $\boldsymbol{\xi}_{n+1}^{\text{net}}$, are the ground truth and prediction latent vectors at time step $n+1$, respectively. During training, the optimizer updates the weights solely of the transformer model, while the encoder and decoder remain frozen. This ensures that the latent space remains consistent with the metriplectic structure imposed during the pretraining phase, allowing the transformer to learn time dependencies without altering the physical meaning of the latent representation (Fig. \ref{Training} (b)).

The internal architecture of the transformer is governed by the attention mechanism, which computes a contextual weighting between latent states, and is based on GPT-2 \cite{OpenAI2018, Wolf2020, Geneva2022}. For each output position $i$, an attention weight $\alpha_{i,j}$ is computed for each input position $j$ via:
\begin{equation}\label{alfa}
\alpha_{i,j}=\text{softmax} \left[ \frac{\boldsymbol{k}(\bs \xi_{j})\boldsymbol{q}(\bs \xi_{i})}{\sqrt{dim(\boldsymbol{k})}}  \right], 
\end{equation}
where $\boldsymbol{k}$, and $\boldsymbol{q}$, are the \textit{key} and \textit{query} vectors, respectively, and $\alpha_{i,j}$ represents the attention score, which modulates the contribution of each value vector $\boldsymbol{v}(\bs \xi_i)$ to the output:
\begin{equation}
\hat{\bs \xi}_i = \bs W^T \sum_{j=n-s}^{n} \alpha_{i,j} \boldsymbol{v}(\bs \xi_j),
\end{equation}
where $\bs W^T$ is a parameter matrix. The \textit{value}, \textit{key}, and \textit{query} vectors are obtained through neural network mappings of the input latent states, denoted as $\mathcal{F}_v(\boldsymbol{\xi}_n)$, $\mathcal{F}_k(\boldsymbol{\xi}_n)$, and $\mathcal{F}_q(\boldsymbol{\xi}_n)$, respectively. The full integration scheme using the Transformer is illustrated in Fig. \ref{Integration} (b).

The Transformer architecture is primarily governed by three hyperparameters: the context length ${\tt n}_{\text{ctx}}=s$, which defines the number of past time steps the model can attend to; the number of attention heads $\tt n_{\text{head}}$, which enables the model to capture multiple correlation patterns in parallel; and the number of layers $\tt n_{\text{layer}}$, which determines the depth of the network and its ability to model complex temporal dependencies. These parameters were chosen to balance expressiveness and computational cost across the considered dynamical systems. For all cases, the activation function used is \texttt{gelu\_new}, commonly used in transformer-based architectures such as GPT-2 \cite{Wolf2020}. After the SPNN is trained, the encoder-decoder is frozen. Then, the transformer is trained with ${\tt n}_{\text{ctx}}=32$, ${\tt n}_{\text{layer}}=6$, and ${\tt n}_{\text{head}}=4$. The optimizer used is Adam, with learning rate $l_r = 1\cdot10^{-3}$, weight decay set to $w_d = 1\cdot10^{-10}$, and a cosine annealing schedule with warm restarts very $1/2$ of the total number of epochs, $n_{\text{epoch}}=850$ \cite{Kingma2014, Loshchilov2017}.

\section{Results}

The performance of the proposed models, the SPNN and the Transformer, have been evaluated for three different physical systems. The first system corresponds to a fluid flow past a cylinder governed by the Navier-Stokes equations, which does not exhibit history-dependent behavior. For the second system, we selected an Oldroyd-B model representing a Couette flow with a viscoelastic fluid. The third system considers a Finitely Extensible Nonlinear Elastic, FENE, fluid, also viscoelastic but characterized by nonlinear behavior of the polymer chains \cite{owens2002computational}. Unlike the first case, both the Oldroyd-B and FENE models incorporate memory effects. Their non-Newtonian response depends on the fluid’s prior strain rate history, as it arises from the presence of dissolved elastic polymers in the solvent.

\subsection{Flow past a cylinder}
In the first case, we consider a Navier-Stokes fluid flow around a cylinder, which does not exhibit any history-dependent behavior. In this case, the state variables are fully characterized by simply storing: 
$$\bs{z}(\bs{x},t):=(u_x,u_y,p),$$
where $u_x$, and $u_y$, denote the velocity components in the $x$, and $y$, directions, respectively, and $p$ is the pressure. 

The ground truth data is obtained by solving the two-dimensional Navier–Stokes equations using the OpenFOAM software \cite{Jasak2007}. The different simulation cases are generated by varying the Reynolds number in the interval $Re \in [100, 750]$, where $Re = u_{\text{in}} d/\nu$, with $u_{\mathrm{in}} = 1$ being the inlet velocity and $d = 2$ the diameter of the cylinder, resulting in $N_{\text{train}}=27$ and $N_{\text{valid}}=6$ separated cases. Each simulation is discretized into $N_T = 400$ snapshots, with a time increment of $\Delta t = 0.5\,\mathrm{s}$. 

The encoder-decoder structure is based on a convolutional architecture that performs sequential spatial reduction as follows:
$$
64 \times 128 \to 32 \times 64 \to 16 \times 32 \to 8 \times 16 \to 4 \times 8,
$$
with increasing feature depths $16 \to 32 \to 64 \to 128$. The resulting representation is flattened and projected to the latent space. The resulting latent vector $\boldsymbol{\xi} \in \mathbb{R}^{n_{\text{embd}}}$ encodes the essential information of the input, where $n_{\text{embd}}$ is a user-defined latent dimmension. The decoder mirrors the encoder, applying a sequence of bilinear upsampling operations interleaved with convolutional layers to reconstruct the original spatial resolution and output the physical variables. 

The activation function used is the leaky-ReLU with a negative slope of 0.1, except for the last layer of both the encoder and decoder, where linear layers are used \cite{Bermejo-Barbanoj2024}. The optimizer used is Adam, with learning rate $l_r = 10^{-3}$ and weight decay set to $w_d = 10^{-7}$ \cite{Kingma2014}. A cosine annealing schedule with warm restarts is applied to the learning rate, with restarts occurring every $1/2$ of the total number of epochs, $n_{\text{epoch}}=850$, in order to promote better exploration of the loss landscape and avoid premature convergence \cite{Loshchilov2017}. The Metriplectic integrator has been defined with $n_{\text{layers}}=6$ hidden layers, with $n_{\text{units}}=34$ hidden units, and the loss weights are $\lambda_{\text{data}}=1.0$, $\lambda_{\text{enc}}=0.1$, and, $\lambda_{\text{deg}}=1\cdot10{-3}$. The dataset is normalized, and Gaussian noise with standard deviation \(1 \cdot 10^{-4}\) is added.

\begin{figure} 
    \centering
    \includegraphics[width=0.87\textwidth]{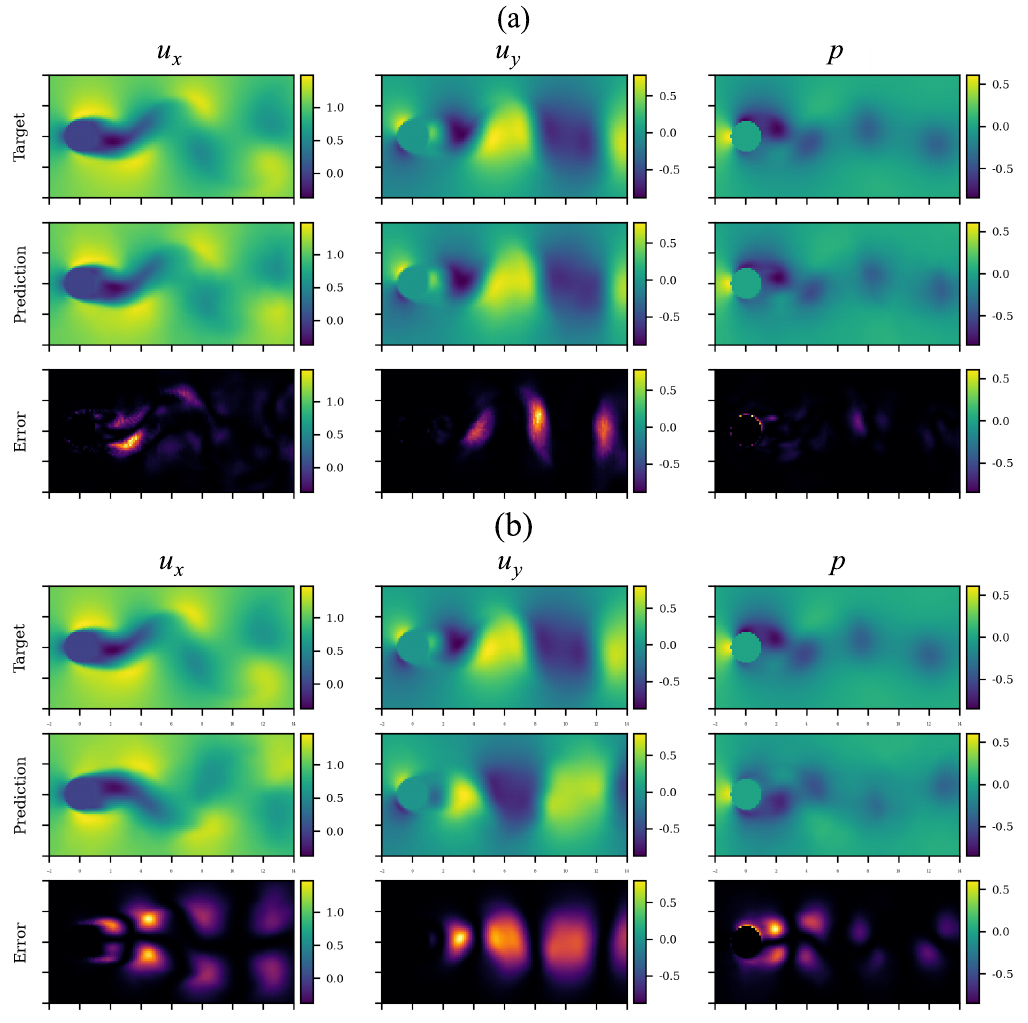}
    \caption{Last snapshot from the rollout results of the flow around a cylinder, in a validation case. (a) The metriplectic neural network shows good agreement with the ground truth. (b) After 398 snapshots, the transformer exhibits a delay with respect to the ground truth.}
    \label{cylinder}
\end{figure}

The results obtained with the two architectures, the SPNN and the Transformer, are compared for the case of flow around a cylinder, using a latent space dimension of $n_{\text{embd}} = 128$. In general, both models exhibit good overall performance. While the SPNN demonstrates high stability throughout the rollout, the Transformer shows a slight delay with respect to the ground truth in the final snapshots of the sequence, see Fig. \ref{cylinder}. 

\begin{figure} 
    \centering
    \includegraphics[width=0.95\textwidth]{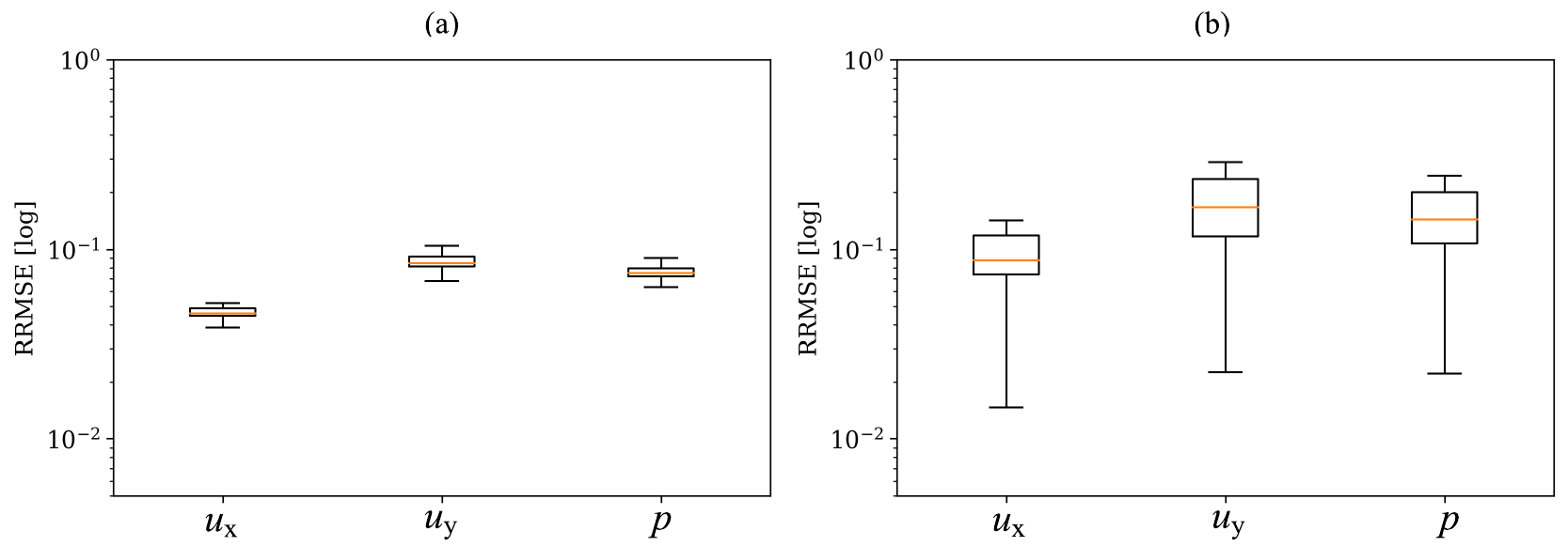}
    \caption{RRMSE for rollout reconstruction on the flow around a cylinder validation dataset: (a) Metriplectic, (b) Transformer. }
    \label{cylinder_RRMSE}
\end{figure}

To quantitatively evaluate the results, the relative root mean square error (RRMSE) for each variable $v_i$ is calculated through
\begin{equation}
    \text{RRMSE}(v_i) = \sqrt{\frac{1}{n_t} \sum_{t=1}^{n_t} \frac{1}{n_p} \sum_{p=1}^{n_p} \frac{\|v_i^{\text{GT}} - v_i^{\text{pred}}\|_2^2}{\|v_i^{\text{GT}}\|_{\infty}^2}},
\end{equation}
where $v_i^{\text{pred}}$ and $v_i^{\text{GT}}$ denote the predicted and ground truth values of the variable $v_i$ at each snapshot $n_t$ and spatial point $n_p$, respectively. 

Errors of the same order of magnitude are observed for all variables in both models (Fig. \ref{cylinder_RRMSE}). However, the Metriplectic model consistently yields lower errors in all variables with reduced variance in the results compared to the Transformer. In this regard, it can be concluded that the metriplectic network is more robust than the transformer. This is clearly related to the ability of the SPNN to ensure compliance with the laws of thermodynamics, something that the transformer, a priori, cannot do. Its comparative advantage, the ability to handle historical information, is of no use in problems governed by the Navier-Stokes equation. However, the transformer provides results that are competitive with SPNN, although of lower accuracy.

We have also trained both models using different latent space dimensions, with $n_{\text{embd}} = 16$, $32$, and $64$. In all cases, the SPNN achieved low RRMSE and successfully reconstructed the rollout sequences, demonstrating its ability to operate effectively under strong compression. In contrast, the performance of the Transformer model deteriorated significantly when reducing the latent space below $n_{\text{embd}} = 128$, and was unable to retain predictive accuracy under such constraints. This highlights the critical role of the embedding dimension in ensuring the performance of attention-based architectures.

\subsection{Viscoelastic fluids}

In viscous fluids, such as Newtonian fluids, the stress tensor is linearly related to the rate of deformation:
\begin{equation}
    \boldsymbol{\tau} = - \mu \dot{\boldsymbol{\gamma}}, 
\end{equation}
where $\mu$ is the dynamic viscosity, and $\dot{\boldsymbol{\gamma}} = \nabla \boldsymbol{v} + (\nabla \boldsymbol{v})^T$ denotes the rate-of-strain tensor. 

However, polymeric or viscoelastic fluids exhibit an additional stress contribution due to the presence of microstructural elements, such as dissolved polymer chains, that store and release elastic energy. In this case, the total extra stress tensor $\boldsymbol{\tau}$ is typically decomposed into two parts: the Newtonian (or solvent) contribution, $\boldsymbol{\tau}_{n}$, and the polymeric (or elastic) contribution, $\boldsymbol{\tau}_{p}$, such that \cite{Laso1993, Bris2009}: 
\begin{equation}
    \boldsymbol{\tau} = \boldsymbol{\tau}_{n} + \boldsymbol{\tau}_{p}.
\end{equation}
The Newtonian part retains the classical viscous form, $\boldsymbol{\tau}_{n} = -\mu_s \dot{\boldsymbol{\gamma}}$, where $\mu_s$ is the solvent viscosity, while the polymeric stress $\boldsymbol{\tau}_{p}$ depends on the specific constitutive model adopted to describe the fluid's memory and elasticity, such as the Oldroyd-B or FENE-P models.

Considering now the Oldroyd-B model, where the viscoelastic stress contribution of the polymer is approximated by a Maxwell model (a 1D spring-dashpot system with viscosity $\mu$ and elastic modulus $E$), the polymeric stress tensor $\boldsymbol{\tau}_{p}$ can be expressed as:
\begin{equation}\label{memory}
\boldsymbol{\tau}_{p} (t) = \int_{-\infty}^{t} \frac{\mu_{p}}{\lambda^{2}} \exp\left( - \frac{t-s}{\lambda} \right) \dot{\boldsymbol{\gamma}}(t,s) , ds,
\end{equation}

where $\mu_{p}$ is the viscosity of the polymer and $\lambda = \mu / E$ denotes the characteristic relaxation time of the system. The term $\frac{\mu_{p}}{\lambda^{2}} \exp\left( - \frac{t-s}{\lambda} \right)$ is commonly referred to as the memory function of the system \cite{Laso1993}.

In the Oldroyd-B framework, the polymeric stress tensor can also be defined from the conformation tensor $\boldsymbol{c} = \langle \boldsymbol{r} \boldsymbol{r} \rangle$, which corresponds to the second moment of the dumbbell end-to-end distance $\bs r$ distribution function. This tensor is not directly measurable and is considered an internal variable of the system. However, using a stochastic approximation, its expected value can be estimated from the relative polymer extension in the flow direction $r_x$ and perpendicular to it $r_y$, as proposed in \cite{Laso1993}:
\begin{equation}
\boldsymbol{\tau}_{p} (t) = \frac{\epsilon}{We} \frac{1}{K} \sum_{k=1}^{K} r_x r_y,
\end{equation}
where $K$ is the number of polymer samples, $\epsilon = \mu_p / \mu$ is the ratio between the polymer and total viscosity, and $We$ is the Weissenberg number, quantifying the ratio between the polymer relaxation time and the characteristic time scale of the fluid.

As stated previously, while the conformation tensor $\boldsymbol{c}$ is not directly observable, the full set of state variables that describe the system should ideally include it. In the case of the Oldroyd-B model, the complete state vector reads: $\boldsymbol{z}(\boldsymbol{x}, t) := (q_x, q_y, u_x, e_i, \bs c), $ where $q_x$ and $q_y$ denote the fluid displacements in the $x$ and $y$ directions, $u_x$ is the flow velocity, $e_i$ is the internal energy, and $\bs c$ represents the conformation tensor characterizing the relative orientation of the polymers (set of variables for Couette flow, 2D, incompressible). 

Nevertheless, given that $\boldsymbol{c}$ is not directly measurable (i.e., a phenomenological variable), the input to the learning model will be restricted to observable instantaneous quantities. Therefore, the state vector for training will be defined as: 
\begin{equation}
\boldsymbol{z}(\boldsymbol{x}, t) := (q_x, q_y, u_x, e_i, \bs \tau),
\end{equation}
where $\bs \tau$ is used as a proxy to $\bs c$, and denotes the total stress tensor, given by $\boldsymbol{\tau} = \boldsymbol{\tau}_{n} + \boldsymbol{\tau}_{p}$. 

The ground truth datasets are obtained with an in-house algorithm based on the CONNFFESSIT technique considering a Couette flow in an Oldroyd-B model fluid \cite{Laso1993, Hernandez2021}. The different simulation cases are generated by varying the Reynolds number in the interval $Re \in [1, 10]$, where $Re = \rho U L/\mu$, with $\rho$ the fluid density, $U = 1$ being the inlet velocity, and $L = 1$ the characteristic length. Then, the training and validation are separated in $N_{\text{train}}=14$ and $N_{\text{valid}}=5$ cases. Each simulation is discretized into $n_p = 200$ section points, and $n_t = 700$ snapshots, with a time increment of $\Delta t = 1e-3\,\mathrm{s}$. 

The encoder-decoder structure is based on a fully connected architecture that first flattens the input state tensor $\boldsymbol{z} \in \mathbb{R}^{d \times N}$, where $d$ denotes the number of physical variables (e.g., velocity components, stress components) and $N$ the number of spatial points. The flattened vector is then projected through a sequence of linear layers as follows:
$$
\mathbb{R}^{d \cdot N} \to \mathbb{R}^{(d \cdot N)/8} \to \underbrace{\mathbb{R}^{\tt n_{\text{hid}}} \to \ldots \to \mathbb{R}^{\tt n_{\text{hid}}}}_{\tt n_{\text{layers}} - 1 \text{ hidden layers}} \to \mathbb{R}^{\tt n_{\text{embd}}},
$$
where $n_{\text{hid}}$ denotes the number of hidden units, and $n_{\text{embd}}$ is the latent space dimension. Additional hidden layers of width $\tt n_{\text{hid}}$ may be inserted between the second and third layer, depending on the configuration parameter $\tt n_{\text{layers}}$. The decoder mirrors this structure in reverse order, and the final output is reshaped into the original spatial format $\mathbb{R}^{d \times N}$.

The activation function used is the leaky-ReLU with a negative slope of 0.1, except for the last layer of both the encoder and decoder, where linear layers are used \cite{Bermejo-Barbanoj2024}. The optimizer used is Adam, with learning rate $l_r = 5\cdot10^{-4}$ and weight decay set to $w_d = 10^{-7}$ \cite{Kingma2014}. A cosine annealing schedule with warm restarts is applied to the learning rate, with restarts occurring every $1/2$ of the total number of epochs, $\tt n_{\text{epoch}}=850$, in order to promote better exploration of the loss landscape and avoid premature convergence \cite{Loshchilov2017}. The encoder-decoder is configured with $\tt n_{\text{layers}}=1$, and $\tt n_{\text{units}}=80$. The Metriplectic integrator has been defined with $\tt n_{\text{layers}}=5$ hidden layers, with $\tt n_{\text{units}}=60$ hidden units, and the loss weights are $\lambda_{\text{data}}=1.0$, $\lambda_{\text{enc}}=0.01$, and, $\lambda_{\text{deg}}=5\cdot10^{-5}$. 

\begin{figure} 
    \centering
    \includegraphics[width=0.8\textwidth]{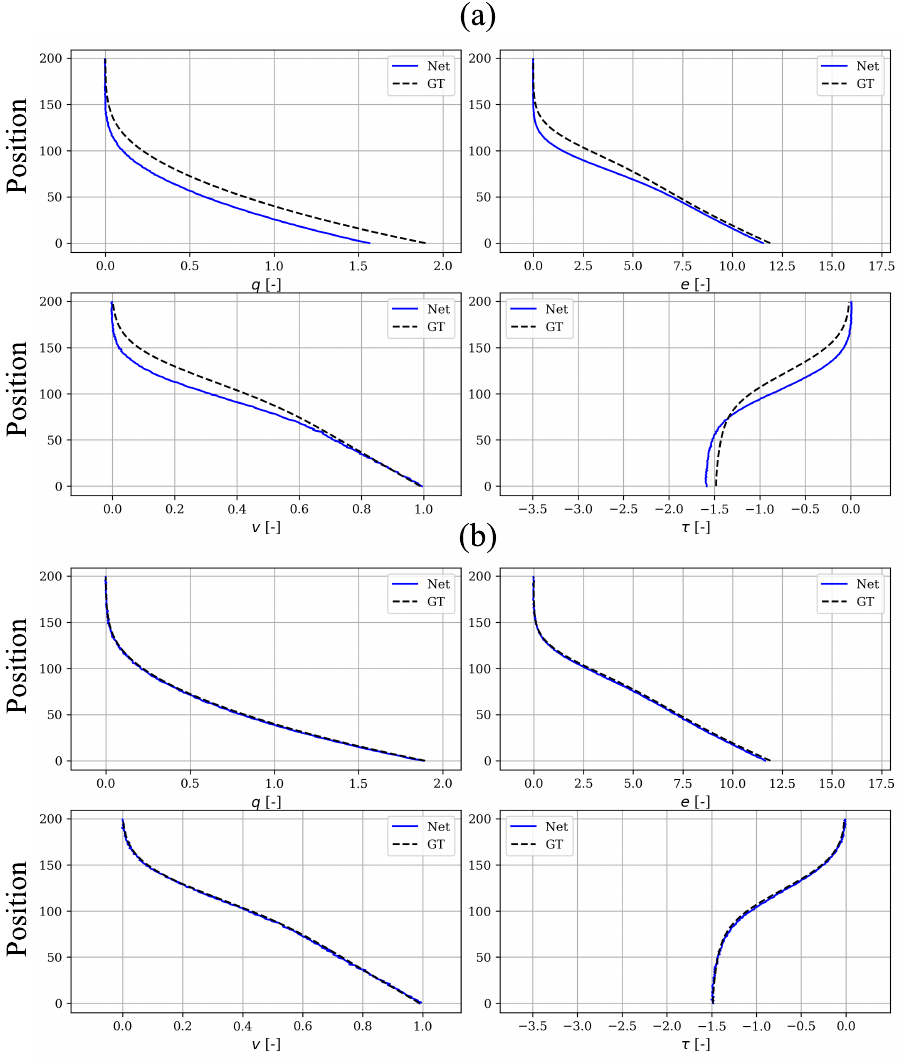}
    \caption{Last snapshot from the Oldroyd-B model rollout of a validation case. The system variables are represented along the 200 points of the discretized section (Y axis). The predictions are represented with a blue solid line, while the ground truth is a black dashed line. The results show a good level of agreement between the ground truth and the predictions obtained by (a) the Metriplectic* neural network and (b) the Transformer model. }
    \label{oldroyd}
\end{figure}

We denote with an asterisk, i.e., Metriplectic$^{*}$ or SPNN$^{*}$ neural network,  when it operates on an incomplete set of variables due to the absence of historical variables. In contrast, the Transformer can access the history through the attention mechanism. As in the previous example, we have compared the  SPNN$^{*}$ with the Transformer using a latent space dimension of $n_{\text{embd}} = 12$. When comparing the reconstruction of the two architectures, SPNN$^{*}$ exhibits a greater discrepancy with respect to the ground truth than the Transformer. In Fig. \ref{oldroyd} only the last snapshot is shown for both cases. However, when examining the results over the entire reconstruction, a large discrepancy is observed for the SPNN$^{*}$ model at intermediate times. This discrepancy is slightly reduced towards the end of the reconstruction, but it does not fully match the ground truth. In this case, access to the history appears to be critical for accurate system reconstruction. 

\begin{figure} 
    \centering
    \includegraphics[width=0.95\textwidth]{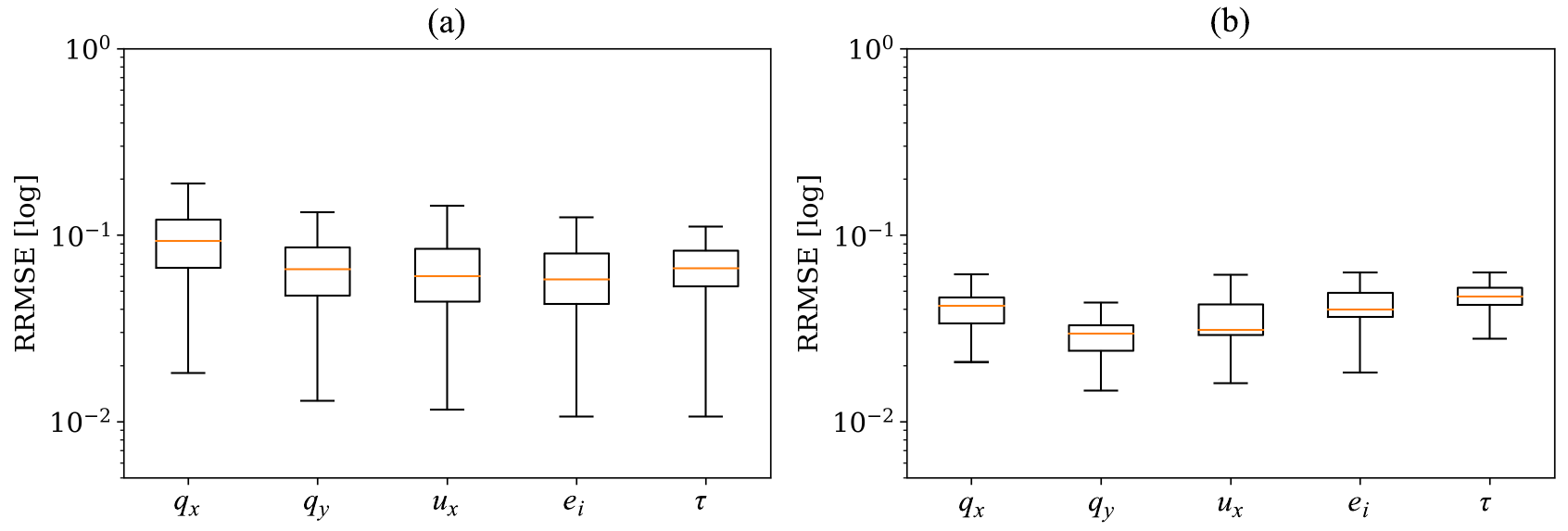}
    \caption{RRMSE for rollout reconstruction on the Oldroyd-B validation dataset: (a) Metriplectic$^{*}$ neural network, (b) Transformer. }
    \label{oldroyd_RRMSE}
\end{figure}

When comparing the RRMSE per variable, a significant difference is observed between the performances of both models, with much better results for the Transformer (Fig. \ref{oldroyd_RRMSE}). In contrast to the previous case, the Transformer exhibits lower variability in the relative error, and the reconstructions are significantly more stable.

\subsection{Non-linear polymeric fluid}

In this section, we consider the same Couette flow configuration described previously, but now the fluid is modeled using a Finitely Extensible Nonlinear Elastic (FENE) model, which introduces strong nonlinearity in the polymeric response \cite{Herrchen1997}. Unlike the Oldroyd-B model, which assumes linear spring forces and allows for infinite extension of the polymer chains, the FENE model incorporates a nonlinear spring force that prevents the dumbbells from exceeding a maximum extensibility. This is achieved by modifying the spring force, $f(\boldsymbol{Q})$, with a finite extensibility correction factor:
\begin{equation}
f(\boldsymbol{Q}) = \frac{1}{1 - |\boldsymbol{Q}|^2 / b},
\end{equation}
where $|\boldsymbol{Q}|$ is the dumbbell extension and $b$ is the finite extensibility parameter. As the polymer approaches its maximum extension, this correction introduces a diverging resistance, effectively constraining the dynamics and introducing additional nonlinearity in the stress response.

The polymeric stress tensor $\boldsymbol{\tau}_p$ is then computed as an ensemble average over the dumbbell configurations as: 
\begin{equation}
\boldsymbol{\tau}_p = \frac{\epsilon}{We} \frac{1}{K} \sum_{k=1}^K \boldsymbol{Q}^{(k)} \otimes f(\boldsymbol{Q}^{(k)}),
\end{equation}
where $\boldsymbol{Q}^{(i)}$ is the configuration of the $k$-th dumbbell, $We$ is the Weissenberg number, and $\epsilon = \mu_p / \mu$ is the ratio between the polymer and total viscosity. The resulting stress tensor is then used as the constitutive input to the macroscopic fluid solver. 

As in the previous case, we define the system state vector as:
\begin{equation}
\boldsymbol{z}(\boldsymbol{x},t) := (q_x, q_y, u_x, e_i, \tau).
\end{equation}
It is important to note that, as in the Oldroyd-B model, this set of observable variables does not provide direct access to the internal microstructural configuration of the fluid. In particular, the polymeric stress tensor $\boldsymbol{\tau}_p$, or equivalently the conformation tensor, is required to fully characterize the dynamics of the system. Therefore, the problem remains a partially observed dynamical system where relevant phenomenological variables must be inferred from the history of the observable quantities.

The ground truth datasets are obtained with an in-house code based upon \cite{Cueto2011, Herrchen1997}, which follows a stochastic formulation of the FENE model using a Monte Carlo approach over ensembles of polymer dumbbells, governed by a Langevin-type equation. The different simulation cases are generated by varying the Reynolds number in the interval $Re \in [1, 10]$, where $Re = \rho U L/\mu$, with $\rho$ the fluid density, $U = 1$ being the inlet velocity, and $L = 1$ the characteristic length. 

The structure of the datasets, including spatial and temporal discretization, the hyperparameter configuration, and the encoder-decoder architecture are totaly equivalent to the Oldroyd-B model. In this example we use a cosine schedule to the learning rate, and the loss weights are $\lambda_{\text{data}}=1.0$, $\lambda_{\text{enc}}=0.1$, and, $\lambda_{\text{deg}}=1\cdot10^{-6}$.

\begin{figure} 
    \centering
    \includegraphics[width=0.8\textwidth]{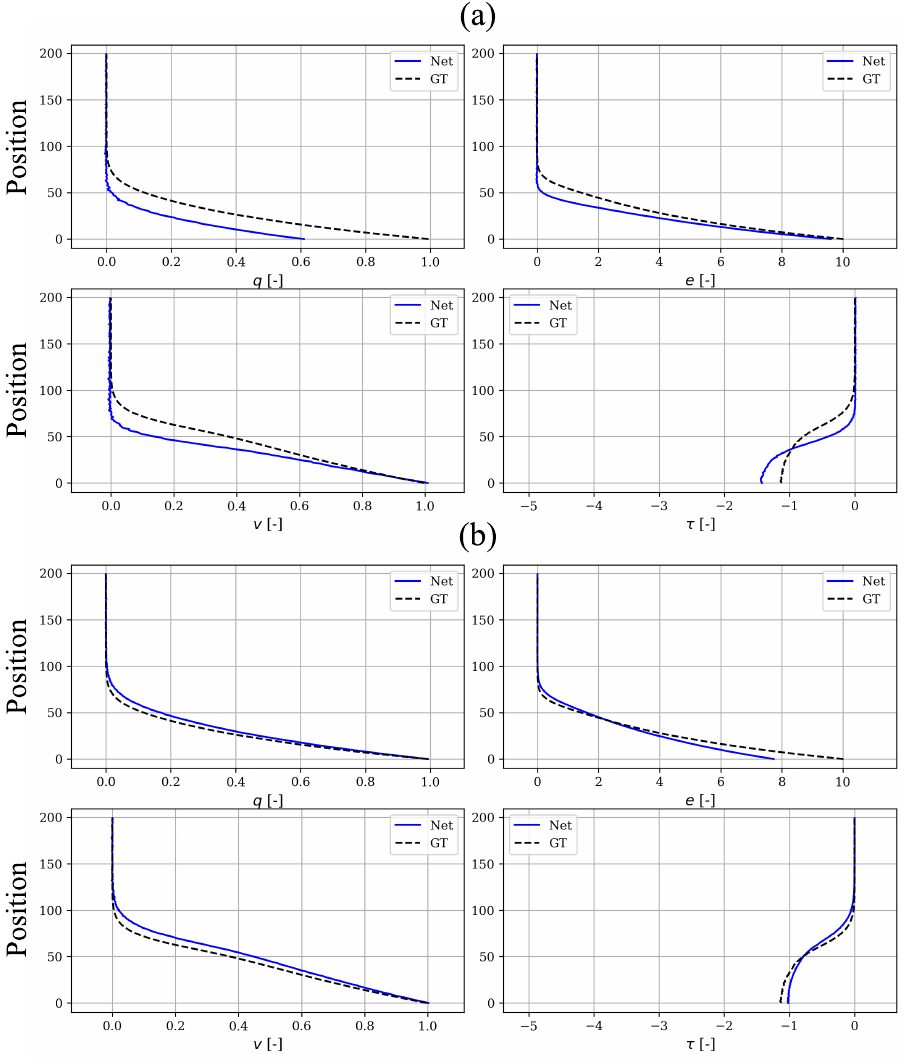}
    \caption{Last snapshot from the FENE model rollout of a validation case. The system variables are represented along the 200 points of the discretized section (Y axis). The predictions are represented with a blue solid line, while the ground truth is a black dashed line. (a) Small deviations from the ground truth are observed with the Metriplectic$^{*}$ neural network, (b) while high reconstruction accuracy is achieved with the Transformer.}
    \label{fene}
\end{figure}

\begin{figure} 
    \centering
    \includegraphics[width=0.95\textwidth]{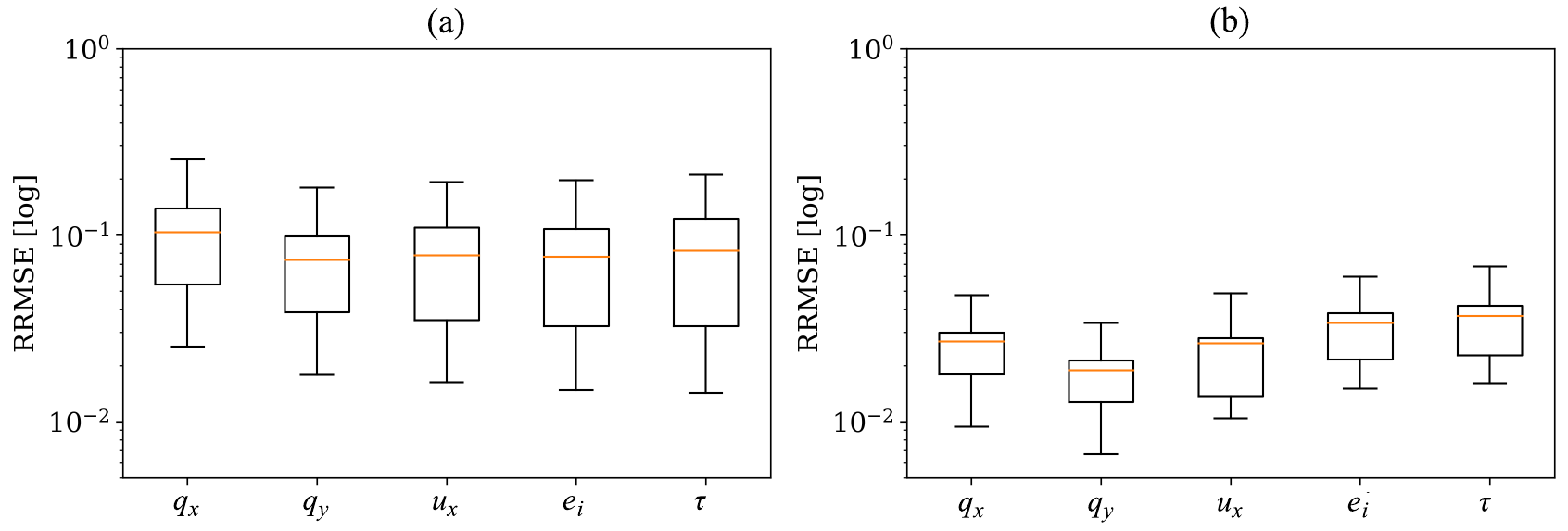}
    \caption{RRMSE for rollout reconstruction on the FENE validation dataset: (a) Metriplectic$^{*}$ neural network, (b) Transformer.}
    \label{fene_RRMSE}
\end{figure}

One of the main challenges in the FENE model lies in its highly nonlinear behavior: while large regions remain static (zero values), the moving fluid exhibits steep gradients. In contrast, the Oldroyd-B model presents much smoother dynamics, with more continuous variable evolution. As a result, the FENE model is considerably more complex due to its stronger nonlinearities. When examining the reconstruction curves for both models, the Transformer appears to fit the Oldroyd-B model more accurately than the FENE model, see Fig. \ref{fene}. However, when evaluating the relative error, lower values are obtained for the FENE model (Fig. \ref{fene_RRMSE}). This is noteworthy given that the use of relative error heavily penalizes zero values in both cases. In the FENE model, more than 50\% of the domain remains at rest (zero values) by the end of the simulation. Despite this, the Transformer achieves lower relative errors for the FENE model than for the Oldroyd-B model.

In contrast, the Metriplectic network shows higher errors for the FENE model, both in terms of relative error and visual agreement with the ground truth. From these observations, we can draw two key conclusions: (1) historical memory is necessary for modeling both systems; and (2) it is even more critical in the FENE model due to its highly nonlinear behavior.

\section{Discussion}


When comparing the overall performance of the three proposed models, a clear relationship is observed between historical memory and reconstruction error for the two architectures under consideration. In the first model—fluid flow around an obstacle, solved with the Navier–Stokes equations—where no historical dependence of the system is present, both architectures exhibit similar reconstruction capabilities, see Fig. \ref{cylinder_RRMSE}, but SPNN networks show slightly lower errors.

We also observe a limitation in the compression applied, which is closely related to the expressiveness of the Transformer. The SPNN, which leverages a specific structure in the data to model the system's evolution, supports high levels of data compression much more robustly. In this regard, while the SPNN achieves predictions with nearly the same relative error even under compressions as high as $\tt n_{\text{embd}} = 16$ (81,072 trainable parameters), the Transformer network reaches relative error values around 30\% (with 20,240 trainable parameters). A high degree of latent space compression significantly reduces the total number of trainable parameters in the Transformer network, thereby limiting its expressive capacity.

\begin{figure} 
    \centering
    \includegraphics[width=0.95\textwidth]{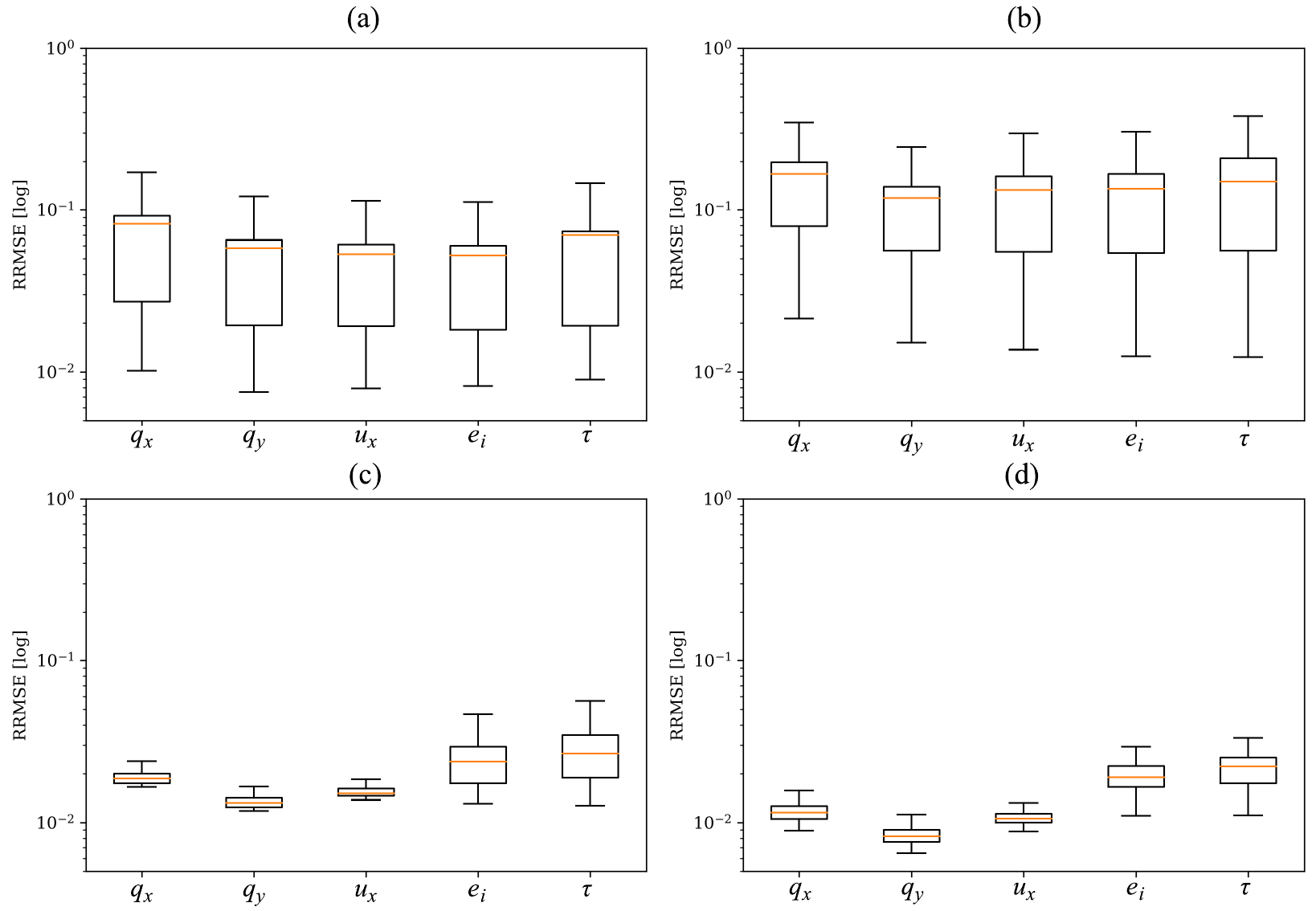}
    \caption{RRMSE for the Oldroyd-B model with different compression levels in the latent space. (a) Metriplectic with ${\tt n}_{\text{embd}}=128$, (b) Metriplectic with ${\tt n}_{\text{embd}}=48$. (c) Transformer with ${\tt n}_{\text{embd}}=128$, (d) Transformer with ${\tt n}_{\text{embd}}=48$. }
    \label{oldroyd_RRMSE_nEmbd}
\end{figure}

The results shown in Fig. \ref{cylinder} and Fig. \ref{cylinder_RRMSE} are compared using a latent space dimension of ${\tt n}_{\text{embd}} = 128$, yielding a similar number of trainable parameters and approximately the same relative error for both architectures (SPNN = 1,224,592; Transformer = 1,208,448). While this setup may appear to be the most fair, it is necessary to consider the following example to observe that the capability—or the main contribution—of the Transformer in this work is not defined by the total number of trainable parameters, but rather by how those parameters are applied. The difference in parameter count arises from the construction of the operators in the formalism, which have dimensions of $\tt n_{\text{embd}} \times \tt n_{\text{embd}}$.

When evaluating the Oldroyd‑B and FENE models using both architectures, we again studied the effect of latent space dimension. In this case, the presented results in Fig. \ref{oldroyd} and Fig. \ref{oldroyd_RRMSE}, and in Fig. \ref{fene} and Fig. \ref{fene_RRMSE}, correspond to a latent space of $\tt n_{\text{embd}} = 12$, with clearly superior performance of the Transformer (11,676 trainable parameters) compared to the SPNN$^{*}$ (169,008 trainable parameters). 

In the previous example, reducing the latent space dimension maintained a stable error for the SPNN model while increasing the error for the Transformer. In the case of the Oldroyd-B model, the roles are reversed. Increasing the latent dimension to $n_{\text{embd}} = 128$ yields minimal improvement in SPNN$^{*}$ results (2,189,960 trainable parameters), but a dramatic reduction in relative error for the Transformer (1,208,448 trainable parameters) (see Fig. \ref{oldroyd_RRMSE_nEmbd} (a) and (c)). For an intermediate case with $n_{\text{embd}} = 48$, the SPNN$^{*}$ network maintains the relative error in the same order of magnitude (447,720 trainable parameters), while the Transformer keeps the relative error low (below 2\% with 172,848 trainable parameters) (see Fig. \ref{oldroyd_RRMSE_nEmbd} (b) and (d)). In this regard, it is clear that network expressiveness alone does not determine superior performance. For the Oldroyd-B model, which shows a history dependence, access to temporal context is key in reconstructing the system’s evolution. In fact, in this case, we observed that with ten times fewer trainable parameters, the Transformer yields clearly superior results compared to the Metriplectic$^*$ network. This is because those parameters are, in essence, attempting to infer a phenomenological variable that is critical for defining the system's evolution. As the Metriplectic model is incomplete (hence the designation SPNN$^{*}$), due to the absence of information on the phenomenological variable, the error of the reconstruction is maintained high. 

The conformation tensor encodes historical information of the system through a memory function, see Eq. (\ref{memory}), and, interestingly, the internal variable $\alpha_{ij}$ is also constructed with a memory function, Eq. (\ref{alfa}) \cite{Laso1993}. While it is not straightforward to establish a direct correlation between both variables, one can reasonably assume that the relevant historical information of the system is encoded in $\alpha_{ij}$. In this sense, we may assert that the historical variable is intrinsically embedded within the Transformer architecture itself. 

This observation opens the door to study systems with historical dependence using models that, through attention mechanisms, may be capable of identifying historical variables based solely on directly measurable quantities. This is a remarkable result, since in practice we typically have access only to instantaneous and observable variables.

\section{Conclusion}

Historical context is essential for constructing predictive models of certain physical systems. In many cases, phenomenological variables are introduced to encode the system’s prior evolution. These internal variables typically reflect hidden structural or microstructural mechanisms—such as fiber orientation in complex fluids or dislocation dynamics in crystalline solids—that are not directly observable at the macroscopic scale \cite{YANG2006}. In this context, the standard strategy is to treat these variables as coarse-grained internal observables, which are not directly accessible at the microscopic level but emerge through macroscopic measurements \cite{Loureno2024}. In this work, we explore an alternative paradigm: the use of attention-based models that implicitly retain historical context, enabling the modeling of history-dependent systems without explicitly introducing internal variables.

The Transformer architecture demonstrates high effectiveness in reconstructing viscoelastic dynamics, significantly outperforming the Metriplectic Neural Network (SPNN$^*$, in the absence of internal variables), while requiring fewer trainable parameters. When compared to the fully metriplectic network (SPNN trained with a complete set of variables) the Transformer still exhibits higher accuracy and efficiency, capturing the history-dependent evolution of the system, using only observable quantities. However, in systems without history dependence, the performance advantage reverses. The metriplectic network achieves better accuracy than the Transformer, even when both architectures have comparable capacity.

These results support the conclusion that Transformer-based models are particularly well-suited for history-dependent systems, where the attention mechanism can effectively prioritize and integrate past states into the current prediction \cite{Liu2025}. Conversely, in systems where the dynamics can be defined solely by instantaneous variables, the inductive bias and resource allocation of the Transformer may be suboptimal, making metriplectic or other physics-informed approaches more appropriate.

In our framework, the encoder-decoder network is first trained to evolve the system via the metriplectic formalism, even in the absence of a complete set of phenomenological variables. This step has a twofold benefit. First, by enforcing a thermodynamically consistent structure, the network tries to encode physically meaningful features, including—where necessary—hidden variables such as the conformation tensor. This improves the expressiveness and relevance of the latent space, making it easier for the Transformer to identify and leverage temporal dependencies within the system  \cite{Hernandez2021a}. Second, the metriplectic prior imposed on the latent representation reduces the complexity of the learning task for the Transformer, which no longer needs to discover the governing structure from scratch but rather focus on modeling the temporal evolution. Finally, while direct interpretability of the latent vectors remains challenging, the previous metriplectic training increases the likelihood that latent variables correspond to physically consistent quantities, enhancing the transparency and scientific relevance of the learned representations \cite{Koune2025, Solera2024}.

\section*{Acknowledgments}

This work was supported by the Spanish Ministry of Science and Innovation, AEI/10.13039/501100011033, through Grant number PID2023-147373OB-I00, and by the Ministry for Digital Transformation and the Civil Service, through the ENIA 2022 Chairs for the creation of university-industry chairs in AI, through Grant TSI-100930-2023-1.

This material is also based upon work supported in part by the Army Research Laboratory and the Army Research Office under contract/grant number W911NF2210271.

This research is also part of the DesCartes programme and is supported by the National
Research Foundation, Prime Minister Office, Singapore under its Campus for Research
Excellence and Technological Enterprise (CREATE) programme.

The authors also acknowledge the support of ESI Group through the chairs at the
University of Zaragoza and at ENSAM Institute of Technology.

\bibliographystyle{unsrt}  

\end{document}